\documentclass[a4paper,aps,prl,preprintnumbers,amsmath,amssymb,twocolumn,longbibliography,superscriptaddress]{revtex4-1} 
\usepackage{amsmath}
\usepackage{amssymb}
\usepackage{verbatim}
\usepackage{graphicx}
\usepackage{color}
\usepackage{xspace}

\begin{document}
 
\title{Fractional angular momentum and anyon statistics of impurities in Laughlin liquids}

\author{Tobias Gra{\ss}}
\affiliation{ICFO-Institut de Ciencies Fotoniques, The Barcelona Institute of Science and Technology, 08860 Castelldefels (Barcelona), Spain}
\author{Bruno Juli\'{a}-D\'{i}az}%
\affiliation{Departament de F\'{i}sica Qu\`{a}ntica i Astrof\'{i}sica, Facultat de F\'{i}sica, Universitat de Barcelona, 08028 Barcelona, Spain.}
\affiliation{Institut de Ci\`{e}ncies del Cosmos, Universitat de Barcelona, ICCUB, Mart\'{i} i Franqu\`{e}s 1, Barcelona 08028, Spain}
\author{Niccol\`o Baldelli}
\affiliation{ICFO-Institut de Ciencies Fotoniques, The Barcelona Institute of Science and Technology, 08860 Castelldefels (Barcelona), Spain}
\author{Utso Bhattacharya}
\affiliation{ICFO-Institut de Ciencies Fotoniques, The Barcelona Institute of Science and Technology, 08860 Castelldefels (Barcelona), Spain}
\author{Maciej Lewenstein}
\affiliation{ICFO-Institut de Ciencies Fotoniques, The Barcelona Institute of Science and Technology, 08860 Castelldefels (Barcelona), Spain}
\affiliation{ICREA, Pg. Lluis Companys 23, 08010 Barcelona, Spain}

\begin{abstract}
 The elementary excitations of a fractional quantum Hall liquid are quasiparticles or quasiholes which are neither bosons nor fermions, but so-called anyons. Here we study impurity particles immersed in a quantum Hall liquid which bind to the quasiholes via repulsive interactions with the liquid. 
 We show that the angular momentum of an impurity is given by the multiple of a fractional ``quantum'' of angular momentum, and can directly be observed from the impurity density.
 In a system with several impurities bound to quasiholes, their total angular momentum interpolates between the values for free fermions and for free bosons. This interpolation is characterized by the fractional statistical parameter of the anyons which is typically defined via their braiding behavior.
\end{abstract}

\maketitle

\textit{Introduction.}
When quasiparticles emerge from strongly correlated quantum matter, their properties can be quite different from those of the matter particles. A paradigm are bulk excitations in fractional quantum Hall (FQH) liquids: The liquid is made of interacting electrons, but its excitations appear as fractional electrons, having fractional charge, fractional angular momentum, and fractional exchange statistics \cite{leinaas77,wilczek82,laughlin83,arovas84}. With this, they are neither bosons nor fermions, but so-called anyons.
To date, the best experimental evidence of fractional quasiparticles is obtained by determining the fractional charge via shot noise measurements \cite{saminadayar97}. Fabry-Perot interferometry \cite{camino07,willett09,ofek10} and beamsplitter experiments \cite{bartolomei20} have provided signatures of fractional statistics. Strong efforts to improve the experimental evidence of anyons concern the implementation of FQH physics in highly controllable quantum systems such as cold atoms \cite{gemelke10,goldman14} or photonic quantum simulators \cite{clark20}.
Light-matter interactions can create and trap fractional quasiparticles in atomic gases \cite{paredes01} or electronic systems \cite{grass18}, and may facilitate braiding operations \cite{paredes01,kapit12,grass14,dutta18}. It has also been suggested to observe the fractional exclusion principle spectroscopically in atomic systems \cite{cooper14}, graphene \cite{papic18}, or magnetic materials \cite{morampudi17}. Moreover, signatures of fractional statistics are carried by the total angular momentum of a fractional quantum Hall system, which can be measured by time-of-flight imaging \cite{umucalilar18}. It has also been proposed to engineer anyonic systems through appropriately defined bath interactions \cite{yakaboylu18,yakaboylu19}. Various works propose to use impurities which bind to fractional quasiparticles \cite{zhang14,zhang15,lundholm16,grusdt16,camacho19}, and which then exhibit features such as fractional relative angular momentum \cite{zhang14}, non-Abelian or Abelian statistics \cite{zhang15,lundholm16}, or quantized transport properties \cite{grusdt16,camacho19}.

Here, we take up the idea of binding impurities to quasiholes in a FQH liquid. First, we consider a single impurity and show that its angular momentum is fractional (in units $\hbar \equiv 1$). Then, by adding more impurities, taken as non-interacting fermions, we observe how the ``anyon sea'' is filled. Specifically, we show that the total angular momentum of the impurities matches neither the value from a fermionic construction, that is by filling the single-particle levels, nor the value of bosonic condensation. Instead, the total angular momentum is reproduced by a linear interpolation between fermionic and bosonic distribution, proportional to $\alpha=1-\nu$. Here, $\nu$ is the filling factor of the FQH liquid, and $\alpha$ equals the anyons' statistical parameter.

 While our results are obtained by numerically solving the underlying quantum Hall model, they can also be understood from fundamental theoretical concepts, and may thus serve as an illustration thereof. In fact, the relation of anyonic physics and fractional quantization of angular momentum dates back to earliest work on the subject: In Wilczek's picture of anyonic statistics \cite{wilczek82}, the fractional behavior emerges through the attachment of matter particles to fluxes, i.e. vortex lines. In Laughlin's wave function for FQH liquids \cite{laughlin83}, the matter particles appear as fluxes seen by the quasihole, and on a mean-field level, this flux attachment re-defines the quasiholes' effective vacuum, i.e., the effective magnetic field seen by the quasiholes. In Ref. \cite{zhang14}, this reasoning has already been employed to explain the fractional relative angular momentum between two anyons, which can be measured via the correlation function of impurities bound to quasiholes. In the present Letter, we demonstrate that the properties of the anyon vacuum and fractional angular momentum can even be probed with a single excitation. The fractionalization of angular momentum can directly be inferred from the density of impurities bound to quasiholes, making it easily accessible in experiment.

The characterization of the single-particle levels provides crucial information which we then use to reveal the anyonic quantum statistics of the impurities. Usually, the statistical behavior of anyons is defined by considering adiabatic braiding operations \cite{arovas84}. In contrast, this Letter examines how the many-body angular momentum of several impurities is composed of the single-particle values, which yields an immediate fingerprint of the distribution function describing the anyon statistics. In accordance with the general expectation \cite{khare-book}, this distribution interpolates between Bose and Fermi statistics, and strikingly, even for very small systems, the statistical parameter obtained from the interpolation agrees almost perfectly with the predictions from an effective impurity Hamiltonian \cite{lundholm16}.

\textit{System.}
The system consists of two types of particles $a$ and $b$: Majority particles ($a$ type) form a FQH liquid with Landau filling fraction $\nu=1/q$, in which impurity particles ($b$ type) are immersed. For simplicity, we assume similar single-particle physics for both species: They have equal mass $M$, are trapped in the $xy$-plane by harmonic potentials of frequencies $\omega_a$ and $\omega_b$, and are brought into the lowest Landau level by a sufficiently strong gauge potential ${\bf A}=\frac{B}{2}(-y,x,0)$. In this gauge, the Fock-Darwin functions, $\varphi_m(z)=(2\pi m! 2^m)^{-1/2} z^m e^{-|z|^2/4}$, are characterized by an angular momentum quantum number $m$. The corresponding single-particle energies are $\epsilon_{m,s}=\hbar m \Omega_s$, with $s \in \{a,b\}$, and $\Omega_s \equiv \sqrt{\omega_B^2+\omega_s^2}-\omega_B$. Here, $\omega_B=eB/M$ is the cyclotron frequency, with $e$ the electric or synthetic charge of the particles. The coordinates $z=(x+iy)/l_B$  are given in units of the oscillator length $l_B=(\hbar/M\Omega_s)^{1/2}$, which depend on the trapping frequency. We assume $\omega_a \approx \omega_b \ll \omega_B$, such that $\Omega_a \approx \Omega_b$, and the length scale $l_B$ takes the same value for both $a$ and $b$.

To make the $a$ particles form a FQH liquid, we consider repulsive interactions. Conveniently, interactions are expressed by Haldane pseudopotentials $U_\ell$, which parametrize the strength of interactions for pairs of particles with fixed relative angular momentum $\ell$  \cite{haldane83}. By truncating the pseudopotential expansion at $\ell=q$ (i.e. by setting $U_\ell =0$ for $\ell \geq q$), we obtain a parent Hamiltonian for the Laughlin liquids at $\nu=1/q$. Its ground state is exactly given by the Laughlin wave function $\Psi_q \sim \prod_{i<j \in a} (z_i-z_j)^q e^{-\sum_i |z_i|^2/4}$, and it has zero interaction energy. The total angular momentum of the Laughlin ground state is $L_q=\frac{q}{2}N_a(N_a-1)$, with $N_a$ the number of $a$ particles. No eigenstates of zero interaction energy are possible for $L_a<L_q$, and within the Hilbert space with $L_a=L_q$, the Laughlin wave function $\Psi_q$ is non-degenerate. Laughlin liquids can be formed either by fermionic or bosonic $a$ particles, depending on whether $q$ is odd or even.

When the angular momentum of the liquid is increased above $L_q$, i.e. for $L_a=L_q+d$ with $d>0$, the liquid can accommodate a characteristic number ${\cal N}_d$ of zero-energy modes. Their wave function is of the form $\Psi_{q,d}^\alpha=f_d^\alpha(\{z_i\}) \Psi_{q}$,  where $f_d^\alpha(\{z_i\})$ is an arbitrary symmetric polynomial of degree $d$. The index $\alpha$ runs from 1 to ${\cal N}_d$, and ${\cal N}_d$ equals the number of partitions of the positive integer $d$. These states describe deformations at the edge, when $d \sim 1$,  but for $d\sim N_a$ they may also describe quasiholes in the bulk. Specifically, the function $\Psi_{q,{\rm qh}} \sim \prod_i (w-z_i) \Psi_q$ describes a quasihole at position $w$. For $w=0$, the factor $\prod_i (w-z_i)$ becomes a symmetric polynomial of degree $d=N_a$, and the state belongs to the manifold of zero-energy solutions at $L_a=L_q+N_a$. 

The $b$ species are taken as non-interacting fermions. To bind to quasiholes of the Laughlin liquid, we consider a sufficiently strong repulsive contact interaction between $a$ and $b$ particles. This interaction allows for exchange of angular momentum between the species, but the joint angular momentum $L=L_a+L_b$ remains a conserved quantity. For the case of a single impurity, the quasihole state $\Psi_{q,{\rm imp}} \sim \prod_i (w-z_i) \Psi_q$ is a state of zero interaction energy, where the dynamical variable $w$ represents the position of the impurity. The interspecies repulsion makes this state non-degenerate at $L=L_q+N_a$, and no zero-energy states exist at $L<L_q+N_a$. Degenerate zero-energy solutions exist at $L=L_q+N_a+d$ with $d>0$, of the form $\Psi_{q,m_1,m_2}^\alpha \sim w^{m_1}f_{m_2}^\alpha (\{z_i\})\prod_i (w-z_i) \Psi_q$, where $m_1$ and $m_2$ are positive integers with $m_1+ m_2=d$. Thus, the number of zero-energy modes at $L=L_q+N_a+d$ is given by ${\cal N}_{d,\rm imp} = \sum_{m_2=0}^d {\cal N}_{m_2}$, see Table~\ref{edgecounting}.

\begin{table}
 \begin{tabular}{|l||c |c |c |c |c |c |c|}
  \hline
  $d$ & 0 & 1 & 2 & 3 & 4 & 5 & 6 \\
  \hline
  ${\cal N}_d$ & 1 & 1 & 2 & 3 & 5 & 7 & 11 \\
  \hline
  ${\cal N}_{d,{\rm imp}}$ & 1 & 2 & 4 & 7& 12 & 19 & 30 \\
  \hline
 \end{tabular}
\caption{\label{edgecounting} Number ${\cal N}_d$ of edge modes in the Laughlin liquid of degree $d$, and number ${\cal N}_{d,{\rm imp}}$ of zero-energy modes of degree $d$ in the presence of an impurity.}
\end{table}

\textit{Results for a single impurity.}
The Laughlin state $\Psi_q$ can be seen as an effective impurity vacuum, and the states $\Psi_{q,{\rm imp}}$ and $\Psi_{q,m_1,m_2}^\alpha$ define the ground state and excited states of a single impurity. These states have total angular momentum $L=L_q+N_a$ and $L=L_q+N_a+m_1+m_2$, but it is not immediately clear how the angular momentum is distributed between the two species. Let $L_b^0$ denote the average angular momentum of the impurity in its ground state, i.e. $L_b^0 \equiv \langle \Psi_{q,{\rm imp}}| \hat L_b | \Psi_{q,{\rm imp}} \rangle$ with $\hat L_b$ the angular momentum operator for the $b$ particle. Naively, one may expect that the angular momentum $L_b^m$ of an impurity in its $m$th excited state, i.e. in $L_b^m \equiv \langle \Psi_{q, m,d-m}^\alpha| \hat L_b | \Psi_{q, m,d-m}^\alpha\rangle$, is given by $L_b=L_b^0+m$. However, as we show below, this is not the case. Instead, the angular momenta of impurity levels differ by multiples of a fractional value, suggesting the interpretation of fractional quantization.

Analytical arguments for this behavior are based on the notion that the impurity at $w$ ``sees'' the majority particles at $z_i$ as fluxes, reducing the effective gauge field for the impurity to $B^* = B- 2\pi \ell_B^2 \rho_a B= B(1-\nu)$, where $\rho_a$ is the density of the majority particles \cite{zhang14}. This leads to an increased magnetic length scale $l_B^* = l_B/ \sqrt{1-\nu}$. Thus, the renormalized wave functions for a single impurity are given by 
\begin{align}
\tilde \varphi_{m}(w) = \sqrt{\frac{(1-\nu)^{m+1}}{2\pi 2^{m} m!}} w^{m} e^{-(1-\nu)|w|^2/4}.
\end{align}
In the limit of $\nu=0$, this wave function is identical to the unrenormalized wave function $\varphi_m(w)$ . The density corresponding to $\tilde \varphi_m$ is given by
\begin{align}
\tilde \rho_{m}(w)&= |\tilde \varphi_m(w)|^2 = \frac{(1-\nu)^{m+1}}{2\pi 2^{m} m!} |w|^{2 m} e^{-(1-\nu)|w|^2/2} \nonumber \\
 &= \sum_{n=0}^\infty \rho_{m+n}(w) \nu^n (1-\nu)^{m+1} \frac{(m+n)!}{m! n!}.
\end{align}
In the second line, we have expanded the renormalized density $\tilde \rho_m$ in terms of unrenormalized densities $\rho_{n+m}=|\varphi_{n+m}|^2$, corresponding to angular momentum  $n+m$. Thus, the average angular momentum $L_b^m$ of an impurity in level $m$ is given by
\begin{align}
\label{qm}
 L_b^m = \sum_{n=0}^\infty (n+m) \nu^n (1-\nu)^{m+1} \frac{(m+n)!}{m! n!} = \frac{m+\nu}{1-\nu}.
\end{align}
In its ground state ($m=0$), the impurity has average angular momentum value $L_b^0=\nu/(1-\nu)$, and exciting the impurity by one unit (from $m$ to $m+1$) changes the average angular momentum by $\Delta L_b = 1/(1-\nu) >1$. The standard deviation is $\delta L_b^m= \sqrt{\nu (m+1)}/(1-\nu)$, so the relative error $\delta L_b^m/L_b^m \rightarrow 0$ for large $m$.

\begin{figure}
\centering
\includegraphics[width=0.49\textwidth, angle=0]{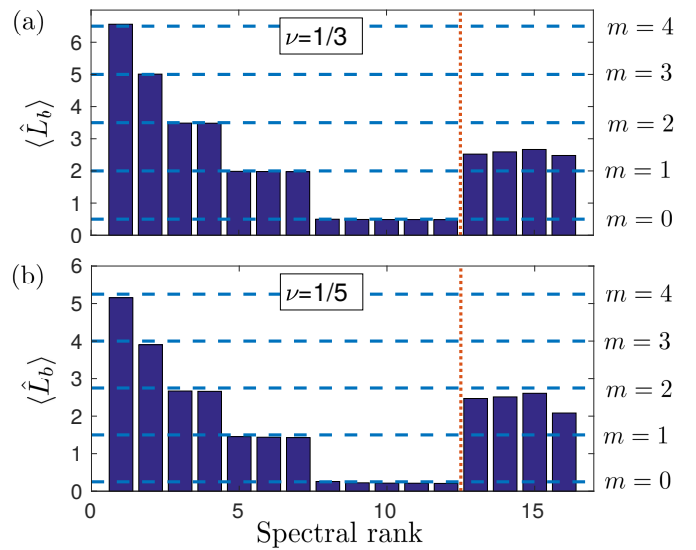}
\caption{\label{fig1} We plot the angular momentum $\langle \hat L_b \rangle$ of an impurity in a Laughlin liquid at (a) $\nu=1/3$ ($N_a=8$ particles at $L=L_3+N_a+4=96$), and at (b) $\nu=1/5$ ($N_a=6$ particles at $L=L_5+N_a+4=85$). The twelve lowest states (on the left to the red-dotted line) are states of zero interaction energy. On average, the impurity takes fractionally quantized values $L_b=\frac{m+\nu}{1-\nu}$ (indicated through the blue-dashed lines).
}
\end{figure}

We have used different methods to verify these results numerically: (i) Applying numerical diagonalization to the pseudopotential Hamiltonian at fixed total angular momentum $L$, the analytical construction of the zero-energy modes can be verified, and in particular the counting of Table~\ref{edgecounting}. We lift the ground state degeneracy ${\cal N}_{d,{\rm imp}}$ at $L=L_q+N_a+d$ by choosing the trap frequency $\omega_a$ slightly larger than $\omega_b$. The states within the quasi-degenerate manifold are then energetically ordered decreasingly with the excitation level $m$ of the impurity: the unique ground state is $\Psi_{q,d,0}$, followed by $\Psi_{q,d-1,1}$, and subsequently two degenerate states $\Psi_{q,d-2,2}^\alpha$, etc. The corresponding impurity angular momentum $\langle \hat L_b \rangle$ is immediately obtained from the numerical solution, and for each $m\leq d$, we find ${\cal N}_{d-m}$ degenerate states, in which the impurity's angular momentum matches very well with the theoretically expected value $ L_b^m= L_b^0 + m \Delta L_b$. This behavior is exemplified in Fig. \ref{fig1} for two cases corresponding to Laughlin filling factors $\nu=1/3$ and $\nu=1/5$. In this example, we have chosen $d=4$ yielding twelve quasi-degenerate states (left of the red-dotted vertical line).

(ii) Eq.~(\ref{qm}) can also be verified by evaluating the impurity angular momentum from the first-quantized wave functions, either by symbolical operations (cf. Ref. \cite{bruno12}), or, attaining much larger system sizes (e.g. $N_a \sim 40$), numerically via Monte Carlo integration method. We have used this method to determine the impurity angular momentum of $\Psi_{q,1,0}$ for $2\leq q \leq 6$, which is accurately given by $L_b^0$.

\begin{figure}
\centering
\includegraphics[width=0.49\textwidth, angle=0]{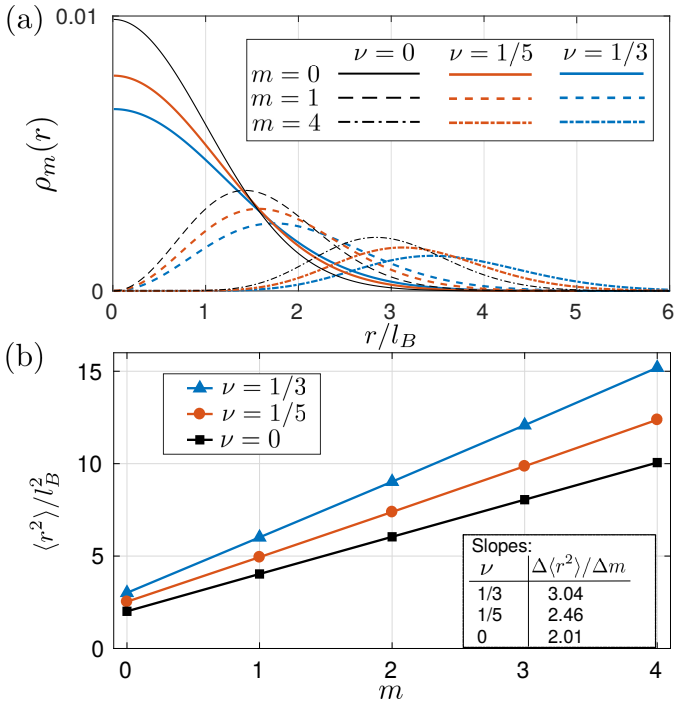}
\caption{\label{fig2} (a) We plot the radial density $\rho_b^{m,q}(|w|)$ of an impurity, which is excited to the $m$th level ($m=0,1,4$), and which is immersed in a FQH liquid at $\nu=1/q$ (for $q=3$ and $q=5$). For concreteness, we have assumed a liquid of $N_a=8$ particles at $L=L_3+N_a+4=96$ for $\nu=1/3$, and a liquid of $N_a=6$ particles at $L=L_5+N_a+4=85$ at $\nu=1/5$, and different $m$ levels correspond to different edge modes. We also plot the density $\rho_m(|w|) = |\varphi_m(w)|^2$ of a single impurity in the absence of a liquid ($\nu=0$). (b) For different levels $m$, we plot the mean square $\langle r^2 \rangle_{m}$ of the radial position of the impurity in the presence of a liquid at $\nu=1/3$, $\nu=1/5$, and in the absence of the liquid. The slope of the linear relation between $m$ and $\langle r^2 \rangle_m$ characterizes the quantization of angular momentum. 
}
\end{figure}

The fractional ``quantization'' of angular momentum is reflected by the impurity density, plotted in Fig.~\ref{fig2}(a). Higher orbitals correspond to larger angular momenta and are characterized by broader density profiles. More quantitatively, there is a linear relation between the mean square of the radial position, $\langle r^2 \rangle$, and the angular momentum $m$. In the absence of a liquid (i.e. for $\nu=0$), we have $\langle r^2 \rangle_m \equiv \int_0^{\infty} dr r^3 |\varphi_m(r)|^2 = 2m+2$. As we find numerically, the slope of this curve changes at finite $\nu$ [see Fig.~\ref{fig2}(b)]. In this case, $\langle r^2 \rangle_{m,q} \equiv \int_0^{\infty} dr r^3 \rho_b^{m,q}(r)$, where the impurity density $\rho_b^{m,q}(r)$, corresponds to a many-body state $\Psi_{q,m,d-m}^{\alpha}$, and is essentially independent from the choice of $d$ and $\alpha$. Specifically, at $\nu=0$, the slope of value 2 corresponds to integer quantization of angular momentum, whereas at $\nu=1/3$ and $\nu=1/5$, the slopes are increased by factors $3/2$ and $5/4$, in full accordance with the expected ``quantization'' of angular momentum.

\textit{Generalization to Moore-Read liquid.} 
The fractionalization of impurity angular momentum, as described by Eq.~(\ref{qm}), does not only apply to impurities in a Laughlin liquid, but also in the non-Abelian Moore-Read liquid incorporating the pairing of particles. Such liquid allows for two types of quasiholes~\cite{wan08}: a ``Laughlin''-like quasiholes, of charge $\nu e$, which is anticorrelated with all liquid particles, and a ``Pfaffian''-like quasihole, of charge $\nu e/2$, which is anticorrelated only with one particle of each pair. By Monte-Carlo integration of their wave functions (see also Supplemental Material \footnote{See Supplemental Material for a discussion of impurities in Moore-Read liquids, long-ranged Hamiltonians, as well as some modifications and limitations of our scheme. The Supplemental Material includes references to \cite{papic11,fey20}.}), we verify that Eq.~(\ref{qm}) holds for an impurity bound to a ``Laughlin''-type quasihole in the Moore-Read liquid. In contrast, for an impurity bound to a ``Pfaffian''-type quasihole, the formula has to be modified by replacing $\nu$ with $\nu/2$, that is, $L_b^m = \frac{2m+\nu}{2-\nu}$. This modification accounts for the fact that the ``Pfaffian'' quasihole only ``sees'' half of the liquid particles.

\begin{figure}
\centering
\includegraphics[width=0.49\textwidth, angle=0]{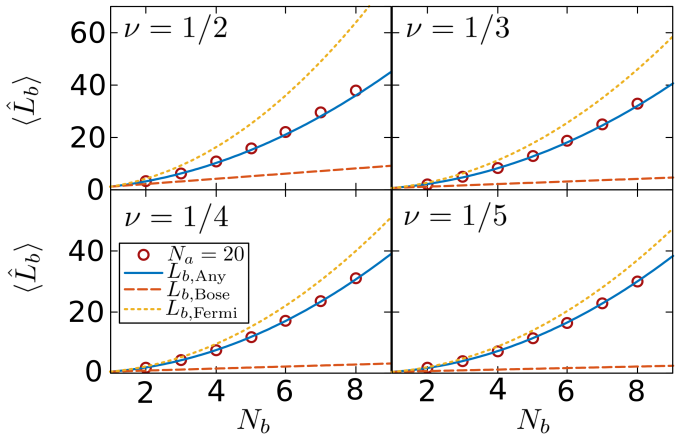}
\caption{\label{Lb} The impurity angular momentum $\langle \hat L_b \rangle$ is plotted as a function of impurity number $N_b$, for different filling factors $\nu$ of the majority liquid in the Laughlin state. The numerical results are obtained from Monte Carlo sampling in the wave function Eq.~(\ref{wavefunc}) for $N_a=20$ majority particles. We also plot $L_{b,{\rm Bose}}(N_b,\nu)$ and $L_{b,{\rm Fermi}}(N_b,\nu)$, the values expected if free bosons or fermions would fill the effective single-particle levels for impurities bound to quasiholes, as well as the anyonic interpolation between both curves, $L_{b,{\rm Any}}(N_b,\nu)$, defined in Eq.~(\ref{interpol}). The numerical data is found to match very well the anyonic prediction.
}
\end{figure}

\textit{Results for multiple impurities.}
Having established the angular momentum levels of a single impurity, Eq.~(\ref{qm}), we now ask how $\langle \hat L_b \rangle$ behaves in the presence of $N_b$ impurities. To obtain states of zero interaction energy, the total angular momentum needs to accommodate the anticorrelations of the majority liquid, the presence of $N_b$ quasiholes, and, for fermionic impurities, a Vandermonde determinant $\prod_{i<j} (w_i-w_j)$. Thus, the zero-energy ground state occurs at $L=L_q+N_b N_a + \frac{1}{2} N_b(N_b-1)$, and its wave function reads:
\begin{align}
\label{wavefunc}
 \Psi_{\rm F,qhs} \sim \left[ \prod_{i<j}^{N_b} (w_i-w_j) \right] \cdot \left[ \prod_{i=1}^{N_a} \prod_{j=1}^{N_b} (z_i-w_j) \right] \Psi_q.
\end{align}
Naively, one might expect that the total angular momentum of the impurities is equal to the value obtained from filling the single-particle levels, $L_{b,{\rm Fermi}}(N_b,\nu) = \sum_{m=0}^{N_b-1} \frac{m+\nu}{1-\nu} = \frac{1}{q-1}[\frac{q}{2}N_b(N_b-1)+N_b]$. However, this expectation is not correct: Fig.~\ref{Lb} shows our numerical results for $\langle \hat L_b \rangle$ as a function of the number $N_b$ of fermionic impurities, interacting with a bosonic or fermionic liquid ($N_a=20$) at different filling factors $\nu$. For comparison, we also plot $L_{b,{\rm Fermi}}(N_b,\nu)$ as well as the angular momentum expected for Bose condensation in the lowest impurity level, $L_{b,{\rm Bose}}(N_b,\nu) = N_b L_b^0 = N_b \frac{\nu}{1-\nu}$. The numerical value is intermediate, $L_{b,{\rm Bose}} < \langle \hat L_b \rangle < L_{b,{\rm Fermi}}$. More precisely, it matches extremely well with the following interpolation formula:
\begin{align}
\label{interpol}
 L_{b,{\rm Any}}(N_b,\nu) = (1-\nu) L_{b,{\rm Fermi}}(N_b,\nu) + \nu L_{b,{\rm Bose}}(N_b,\nu).
\end{align}
This formula suggests that the statistical parameter $\alpha$, which interpolates from Bose statistics ($\alpha=0$) to Fermi statistics ($\alpha=1$), is given by $\alpha=1-\nu$. This is in agreement with the effective Hamiltonian derived in Ref.~\cite{lundholm16} for impurities coupled to fractional quasiholes (see also Refs.~\cite{rougerie16,lundholm17}), and with the general expectation for a Laughlin quasihole ($\alpha=-\nu$) bound to a fermion ($\alpha=1$). Importantly, we note that similar results as shown in Fig.~\ref{Lb} (with $N_a=20$) can already be obtained for extremely small Laughlin liquids ($N_a<10$), enabling the detection of anyon statistics in microscopic quantum simulators.

\textit{Summary and Outlook.}
We have shown that (i) the effective single-particle states for impurities bound to anyons can be characterized by their fractional angular momentum, and (ii) the filling of these levels is governed by fractional statistics. Our findings provide a way to detect anyonic properties without braiding via the density of impurity particles. This eases anyon detection, possibly also compared to existing schemes based on local density of state measurements \cite{cooper14,papic18}, pair-correlation function of two impurities \cite{zhang14}, or liquid density \cite{umucalilar18}. A key difference of our approach to other proposals involving impurities \cite{papic18, zhang14} is the fact that it keeps all impurity particles fully dynamical. This realizes a non-interacting gas of anyons, and an anyonic distribution function governs the impurity degrees of freedom. With this, the setup is also suited to study, in future work, the intricate thermodynamics of anyon gases.

The implementation of our ideas is possible either in microscopic quantum simulators using atoms or photons \cite{gemelke10,clark20}, or in macroscopic electronic samples with optically created impurities such as excitons or trions. Signatures of excitons bound to fractional quasiparticles have been reported in \cite{byszewski06}, and the exciton density can be detected via scanning-transmission-electron microscopy \cite{nerl17}. Additional information covering long-range Hamiltonians is presented in the Supplemental Material. 
Future work shall explore the potential of our scheme for detecting non-Abelian anyons, and for studying thermodynamics of anyons, including interacting anyons which might themselves form FQH liquids.

 \begin{acknowledgments}
  We acknowledge discussions with Axel Pelster, Klaus Sengstock, and Christof Weitenberg. 
 T.G. acknowledges financial support from a fellowship granted by “la Caixa” Foundation (ID 100010434, fellowship code LCF/BQ/PI19/11690013).
 T.G., N.B., U.B., and M.L. acknowledge funding from 
 ERC AdG NOQIA, Spanish Ministry MINECO and State
Research Agency AEI (FIDEUA PID2019-106901GB-I00/10.13039 / 501100011033,
SEVERO OCHOA No. SEV-2015-0522 and CEX2019-000910-S, FPI), European Social
Fund, Fundaci\'o Cellex, Fundaci\'o Mir-Puig, Generalitat de Catalunya (AGAUR Grant No. 2017 SGR 1341, CERCA program, QuantumCAT U16-011424, co-funded by ERDF
Operational Program of Catalonia 2014-2020), MINECO-EU QUANTERA MAQS (funded
by State Research Agency (AEI) PCI2019-111828-2/10.13039/501100011033), EU Horizon
2020 FET-OPEN OPTOLogic (Grant No 899794), and the National Science Centre, Poland-
Symfonia Grant No. 2016/20/W/ST4/00314.
 B. J.-D. acknowledges funding from Ministerio de Economia y Competitividad Grant No FIS2017-87534-P.
 N.B. acknowledges support from a "la Caixa” Foundation (ID 100010434) fellowship. The fellowship code is  LCF/BQ/DI20/11780033.
 \end{acknowledgments}
 

\end{document}


\title{Supplemental Material: Fractional angular momentum and anyon statistics of impurities in Laughlin liquids}

\author{Tobias Gra{\ss}}
\affiliation{ICFO-Institut de Ciencies Fotoniques, The Barcelona Institute of Science and Technology, 08860 Castelldefels (Barcelona), Spain}
\author{Bruno Juli\'{a}-D\'{i}az}%
\affiliation{Departament de F\'{i}sica Qu\`{a}ntica i Astrof\'{i}sica, Facultat de F\'{i}sica, Universitat de Barcelona, 08028 Barcelona, Spain.}
\affiliation{Institut de Ci\`{e}ncies del Cosmos, Universitat de Barcelona, ICCUB, Mart\'{i} i Franqu\`{e}s 1, Barcelona 08028, Spain}
\author{Niccol\`o Baldelli}
\affiliation{ICFO-Institut de Ciencies Fotoniques, The Barcelona Institute of Science and Technology, 08860 Castelldefels (Barcelona), Spain}
\author{Utso Bhattacharya}
\affiliation{ICFO-Institut de Ciencies Fotoniques, The Barcelona Institute of Science and Technology, 08860 Castelldefels (Barcelona), Spain}
\author{Maciej Lewenstein}
\affiliation{ICFO-Institut de Ciencies Fotoniques, The Barcelona Institute of Science and Technology, 08860 Castelldefels (Barcelona), Spain}
\affiliation{ICREA, Pg. Lluis Companys 23, 08010 Barcelona, Spain}

\begin{abstract}
 In the main text, we have considered analytic Laughlin-type wave functions with quasiholes which bind to impurity particles. This construction provides an exact solution for a system in which the liquid interaction is given by the parent Hamiltonian for the Laughlin state, and the interaction between liquid and impurities is described by a contact potential. 
 In this Supplemental Material, we extend our studies of the main text to account for various deviations from this ideal scenario: (i) We consider the effect of an increased  interaction range, both within the liquid and between liquid and impurity. (ii) We consider fractional quantum Hall liquids beyond the Laughlin paradigm, specifically the case of a Moore-Read liquid with Pfaffian pairing. (iii) We discuss the binding of impurities to quasiparticles instead of quasiholes. (iv) Finally, we also discuss limitations of our scheme, specifically the breakdown at integer filling and for highly excited impurities.
\end{abstract}

\maketitle

\section{ Increasing the range of interactions}
The Laughlin wave function at filling $\nu=1/q$ is an exact solution to a system of particles in the lowest Landau level which interact via Haldane pseudopotentials $V_\ell$ with $\ell<q$. Strikingly, Laughlin liquids are also formed in systems with long-range interactions. Specifically, the $\nu=1/3$ state turns out to be a strongly gapped fractional quantum Hall phase described by the Laughlin wave function for a variety of electronic systems (semiconductors, graphene, etc.). Electrons in these systems interact via a screened or unscreened Coulomb potential. Other potentials which support the 1/3 Laughlin state are systems with dipolar potentials (e.g. dipolar atoms, Rydberg atoms). Here we address the question whether such long-ranged systems exhibit the same angular momentum behavior for impurities bound to quasiholes which was discussed in the main text for the pseudopotential model. Specifically, we will look at Coulombic systems (screened or unscreened) described by the potential \cite{papic11}:
\begin{align}
\label{VC}
 V(r) = \frac{e^2}{\epsilon r} + \alpha\frac{e^2}{\epsilon\sqrt{r^2+d^2}},
\end{align}
where the first term accounts for the unscreened Coulomb interactions in a medium of dielectric constant $\epsilon$, while the second term provides a potential screening via a dielectric plate at distance $d/2$ with dielectric constant $\epsilon'$, and $\alpha = \frac{\epsilon-\epsilon'}{\epsilon+\epsilon'}$.

First, we consider the case of unscreened Coulomb interactions within the liquid, and a rigid-body impurity-liquid interaction of strength $V_0$ (as it may for instance apply to an impurity given by a charge-neutral exciton without dipole moment, cf. Ref. \cite{fey20}). In contrast to our pseudopotential model in the main text, the liquid now possesses a finite amount of interaction energy even in its Laughlin-like ground state, which leads to the following competition: When the angular momentum of the liquid is increased to accomodate a quasihole, the system may either form a quasihole and in this way reduce the interactions between liquid and impurity, or it may reduce interaction energy of the liquid via high-order edge deformations. From this perspective, it is clear that the strength $V_0$ of the liquid-impurity interactions will determine the amount to which each of these two mechanisms applies. Accordingly, considering the average angular momentum $\langle L_b \rangle$ of an impurity in its ground state ($m=0$) as a function of $V_0$, we find significant deviations from the expectation $L_b^0 = \frac{1}{2}$ (for $\nu=\frac{1}{3}$), but these deviations decrease, when $V_0$ is increased. For instance, in a system of five electrons $\langle L_b \rangle=0.28$ when $V_0=V_1$, but converges to the value $\langle L_b \rangle =0.39$, when $V_0 \gtrsim 10 V_1$, with $V_1$ being the $\ell=1$ pseudopotential of the Coulomb interactions given by $V_{\ell}=\Gamma(\ell+\frac{1}{2})/2\Gamma(\ell+1)$ (in units $e^2/l_B \epsilon$). These numbers further improve when the system size is increased: For seven and eight electrons, we have obtained values $\langle L_b \rangle = 0.49$ and  $\langle L_b \rangle = 0.52$

This suggests that our findings from the main text  accurately describe also systems with unscreened Coulomb interaction in the liquid and contact interaction between liquid and impurity. However, the condition $V_0 \gg V_1$ is rather restrictive. In the case of exciton-electron interactions, $V_0$ is determined by the binding energy of a trion \cite{fey20}.
Let us therefore also look at systems with screened Coulomb potentials. For concreteness, we choose $\alpha=-1$ and $d=2\sqrt{2} l_B$ in Eq.~(\ref{VC}). This choice dramatically relaxes the requirements on the strength of $V_0$. More importantly, we may now allow for liquid-impurity interactions and liquid-liquid interactions to be given by the same long-ranged potential $V(r)$. Such scenario could apply to impurities given by charged excitons (trions) or electronic impurities (e.g. with spin polarization opposite to the fractional quantum Hall liquid). In this case, $ \langle L_b \rangle $ takes the value 0.5 in a system for seven electrons.

\begin{figure}[t]
    \centering
    \includegraphics[width=0.48\textwidth]{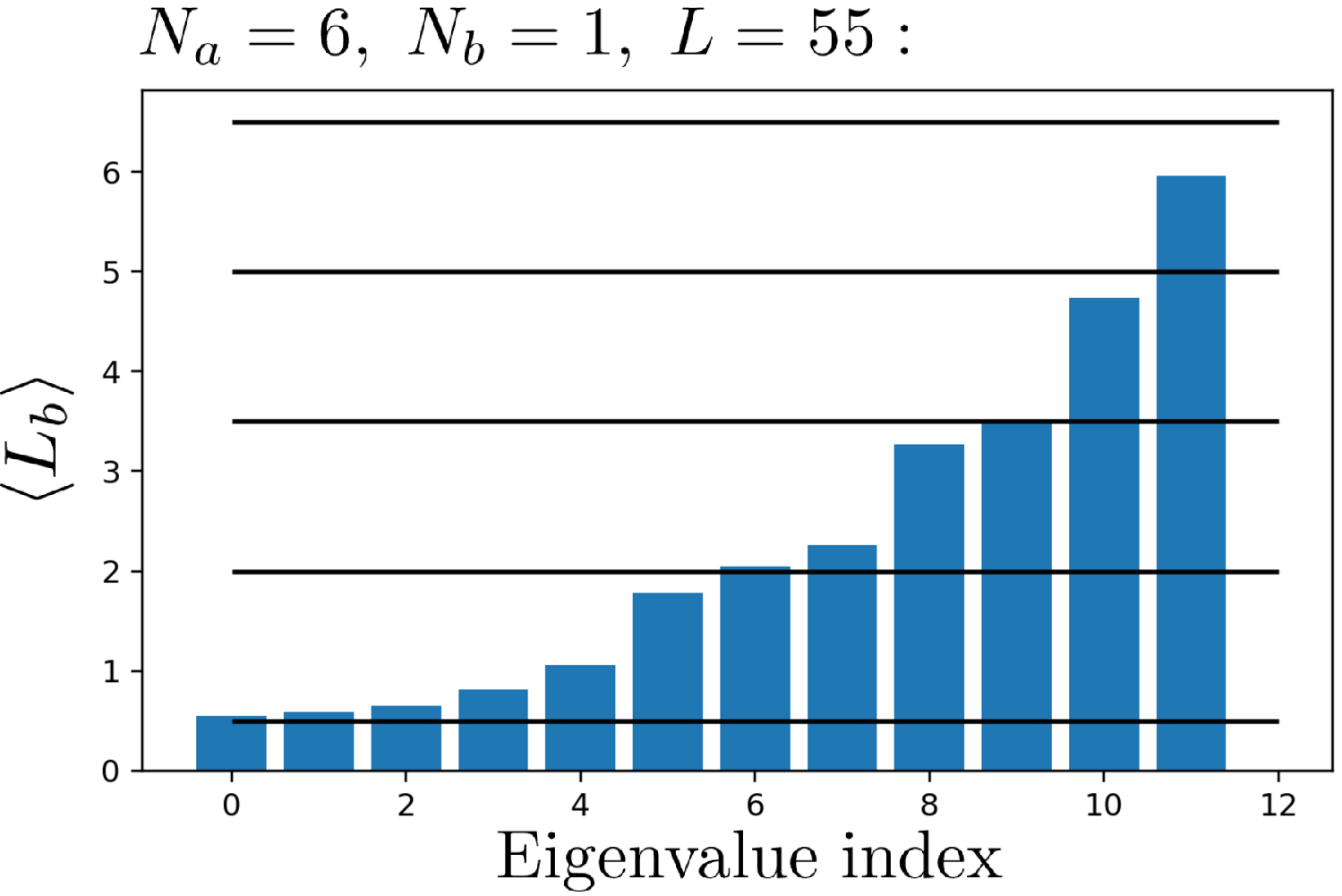}
    \caption{Angular momentum of one impurity interacting via contact interaction with a screened Coulomb liquid of $N_a=6$ electrons, for the twelve quasi-degenerate ground states which exist at total angular momentum $L=L_{\rm Laughlin} + N_a + 4$.
    \label{twelve_levels}}
\end{figure}

Let us further investigate the behavior of impurities in a screened liquid. For concreteness, we will now consider only the case of short-range impurity-liquid interactions, corresponding to charge-neutral impurities. In analogy to Fig. 1 in the main text, Fig.~\ref{twelve_levels} of this Supplemental Material plots the angular momentum of one impurity in the twelve quasi-degenerate ground states which occur when the total angular momentum is four units above the value which corresponds to a unique ground state, $L=L_{\rm Laughlin}+N_a+4$. This twelve-fold manifold is expected to contain five states with $\langle L_b \rangle =0.5$, three states with $\langle L_b \rangle =2$, two states with $\langle L_b \rangle =3.5$, one state at $ \langle L_b \rangle = 5$, and one state at $\langle L_b \rangle =6.5$. Most significant deviations from this ``quantization'' are seen for the latter states at largest $\langle L_b \rangle$, where $\langle L_b \rangle$ remains significantly below the expected values. For the other states at smaller $\langle L_b \rangle$, in contrast, we observe the tendency to exhibit values slightly above the expected values. As a reason for this behavior, we identify the fact that the long-ranged potential lifts the degeneracy of the edge excitations, and thus splits the degeneracy of this low-energy manifold. Therefore, the infinitesimal mismatch between trapping frequency for impurity and liquid particles, which in the totally degenerate case was suited to separate states according to their impurity angular momentum, does not produce the desired splitting anymore. Instead, different excitation levels of the impurity are now mixed. 

The low-energy manifold as a whole, however, is unchanged. Accordingly, if we average $\langle L_b \rangle$ over the full low-energy manifold, we obtain exactly the value (2.25) as obtained by averaging over the twelve ground states in the degenerated model. Therefore, although the degeneracy lifting makes it more difficult or even impossible to determine the ``quantized'' angular momentum values from excited eigenstates, it is not harmful to our actual case of interest, which is the angular momentum of (multiple) impurities in a unique ground state. To this end, we assume that the impurity levels in the long-ranged liquid are still characterized by the same ``quantization'' scheme, i.e. $L_b^m=L_b^0 + \frac{3}{2}m$. This then also defines the value $L_{b,{\rm Any}}$ for anyonic filling of these levels, as given by Eq. (5) of the main text. Accordingly, anyonic filling of the impurity levels should be reflected by $\langle L_b \rangle=L_{b,{\rm Any}}  =2$ for two impurities, or $\langle L_b \rangle = L_{b,{\rm Any}}= 4.5$ for three impurities. Strikingly, our numerical results for the (unique) ground state of two or three impurities at total angular momentum $L=L_{\rm Laughlin} + N_b N_a + N_b(N_b-1)/2$ exhibit values which are very close to the expected value $L_{b,{\rm Any}}$.  The precise numbers are given in Table~\ref{Lb}.

\begin{table}
 \begin{tabular}{|l| c |c |c | c |}
  \hline
  $N_b$  & $N_a$ & $L$ & $L_{b,{\rm Any}}$ & $\langle L_b \rangle$\\
  \hline
 2  & 4 & 27 & 2 & \bf 2.04 \\
 2  & 5 & 41 & 2 & \bf 1.87 \\ 
 2  & 6 & 58 & 2 & \bf 1.91 \\
  \hline
 3 & 6 & 48 & 4.5 & \bf 4.41 \\
 3 & 6 & 66 & 4.5 & \bf 4.41 \\
  \hline
\end{tabular}
\caption{\label{Lb} Filling the anyon sea: $N_b$ impurities interact (via contact interaction, $V_0=1$ with $N_a$ electrons in a screened Coulomb liquid ($\alpha=-1$ and $d=2\sqrt{2} l_B$ in Eq.~\ref{VC}), at Landau filling $\nu=1/3$. The total angular momentum $L$ is chosen such that there is a unique ground state in which the impurities can bind to $N_b$ quasiholes. The angular momentum $\langle L_b \rangle$ of the impurities matches well well with $L_{b,{\rm Any}}$, the expected value for non-interacting anyons defined by Eq.~ (5) in the main text.}
\end{table}

In summary, these results demonstrate that our scheme of characterizing anyons via the angular momentum of impurities bound to the anyons potentially applies also to systems with long-range interactions, provided a sufficiently strong screening of the liquid, and/or sufficiently strong liquid-impurity interactions. 

\section{ Moore-Read liquids}
Our scheme can also be applied to fractional quantum Hall liquids which are not of the Laughlin type. Here, we consider a non-Abelian liquid described by the Moore-Read wave function:
\begin{equation}
    \psi_{\rm MR}(z)=\text{Pf}\left(\frac{1}{z_i-z_j}\right)\prod_{i<j}(z_i-z_j)^q,
\end{equation}
where $\rm Pf$ denotes the Pfaffian. This state describes fermions (bosons) for $q$ even (odd). In presence of $N_a$ particles this state has an angular momentum $L_{\rm MR}=\frac{q}{2}N_a(N_a-1)-\frac{N_a}{2}$.

There are two different types of quasiholes which can be pierced into such liquid: (i) a ``full'' quasi-hole which is in close analogy to the quasihole of the Laughlin liquid, with charge $e/q$ and requiring a total angular momentum $L=L_{\rm MR}+N_a$; (ii) a ``half'' quasi-hole which only repels half of the liquid particles, with charge $e/2q$ and at $L=L_{\rm MR}+\frac{N_a}{2} = \frac{q}{2}N_a(N_a-1)$. The wave function of the full quasi-hole reads
\begin{equation}
    \psi_{\rm qh1}(z,w)=\prod_k(z_k-w)\text{Pf}\left(\frac{1}{z_i-z_j}\right)\prod_{i<j}(z_i-z_j)^q,
    \label{holemr}
\end{equation}
whereas the ``half'' quasi-hole is described by
\begin{equation}
    \psi_{\rm qh2}(z,w)=\text{Pf}\left(\frac{(z_i-w)+(z_j-w)}{z_i-z_j}\right)\prod_{i<j}(z_i-z_j)^q.
    \label{wf}
\end{equation}
As in the main text we assume that these quasiholes are formed due to a repulsive impurity which binds to the quasihole (i.e. $w$ becomes the impurity degree of freedom). In Ref.~\cite{wan08} it has been shown that a Gaussian potential is suited to produce either of the two types of quasiholes, depending on the strength of the repulsion. 

Now we address the question of how the impurity angular momentum behaves when the impurity in either of these two states, $ \psi_{\rm qh1}(z,w)$ or $ \psi_{\rm qh2}(z,w)$, is excited to the $m$th level, that is, by multiplying the respective many-body wave function with a prefactor $w^m$. 

\begin{figure*}[t]
    \centering
    \includegraphics[width=0.48\textwidth]{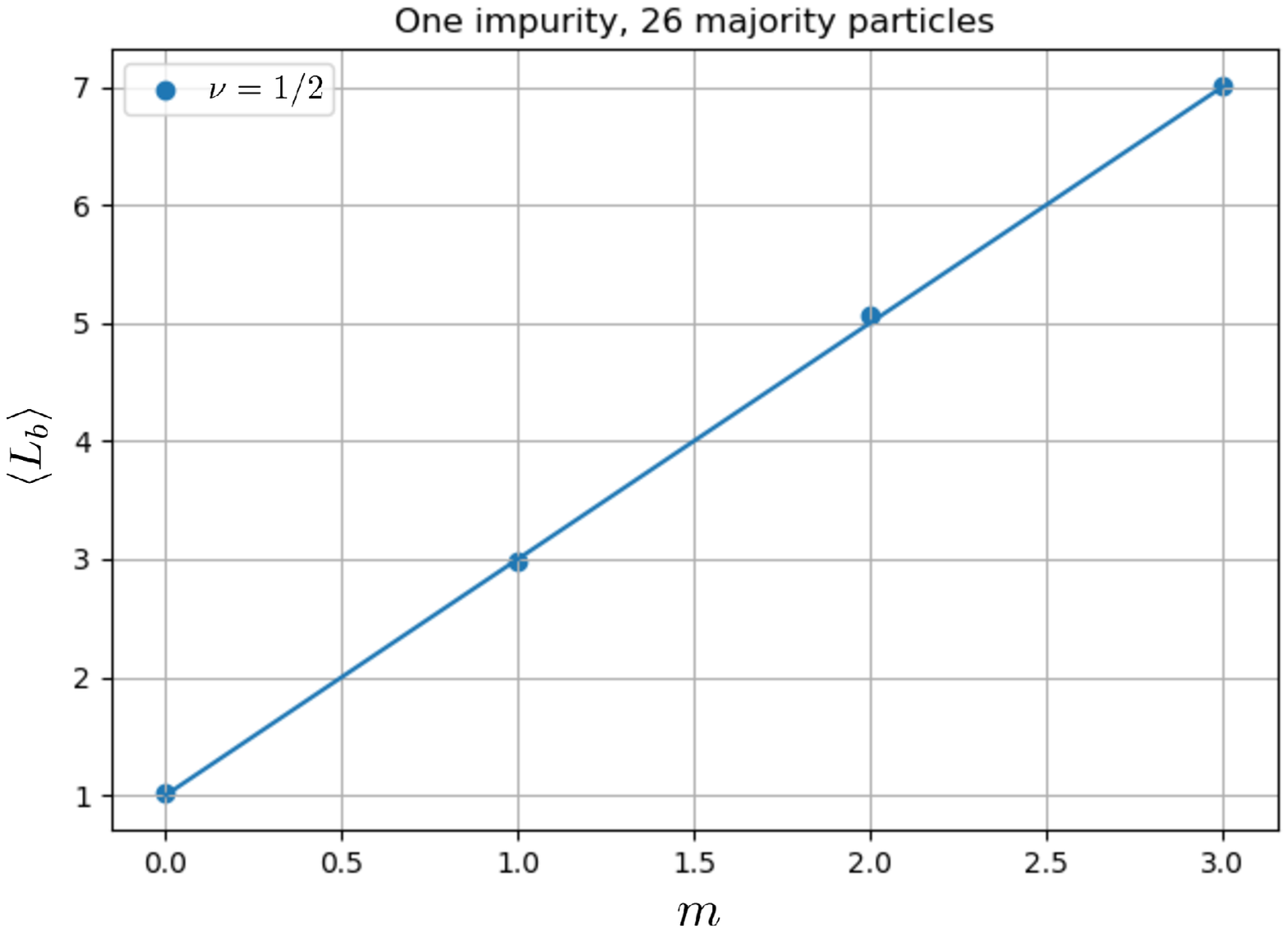} 
    \includegraphics[width=0.48\textwidth]{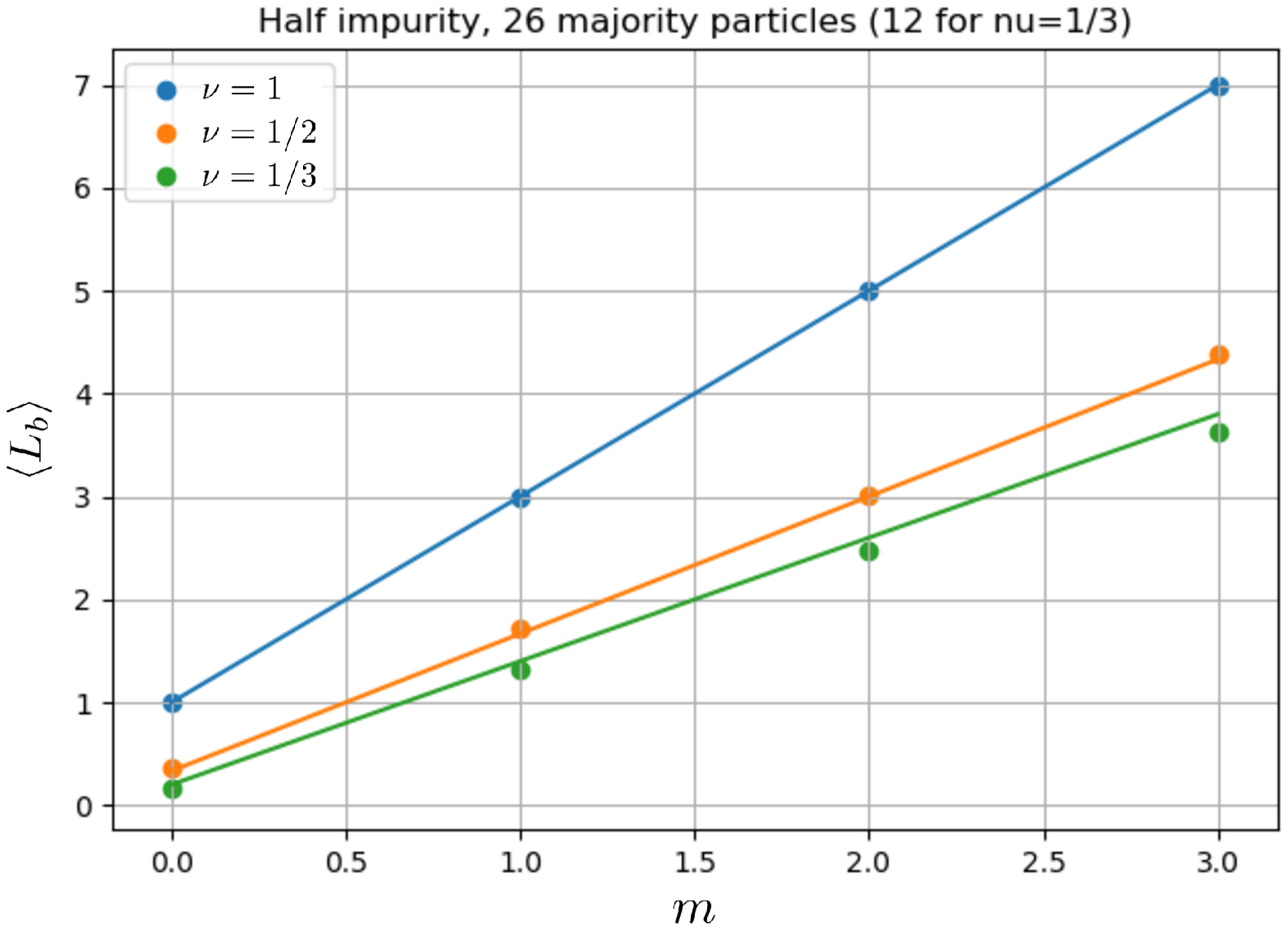}
    \caption{Left: Average angular momentum of one impurity bound to a full quasi-hole in a Moore-Read liquid ($\nu=1/2$), as a function of the excitation index $m$. The solid line represents the expectation from the mean-field formula, Eq.~(3) in the main text. Right: Average angular momentum of one impurity bound to a half quasi-hole in a Moore-Read liquid (different fillings), as a function of the excitation index $m$. The solid line represents the expectation from the mean-field formula, Eq.~(3) in the main text, after replacing $\nu$ by $\nu/2$.}
    \label{MR}
\end{figure*}

The results are shown in Fig.~\ref{MR}. Not surprisingly, the impurity bound to a full quasihole in the $\nu=1/2$ Moore-Read liquid behaves in exactly the same way as an impurity in the $\nu=1/2$ Laughlin liquid, e.g. $L_b^m = \frac{m+\nu}{1-\nu}$. However, for the bosonic Moore-Read liquid at $\nu=1$ this mean-field formula is not applicable, and the impurity angular momentum is found to become a quantity which scales extensively with the number of liquid particles. We note that this seems to be a generic behavior at $\nu=1$, which is further discussed for the case of an integer quantum Hall system in the section below.

For the angular momentum values of the impurity bound to a half quasi-hole, we find that the mean-field formula has to be modified by replacing $\nu$ with $\nu/2$:
\begin{align}
 L_b^m = \frac{2m+\nu}{2-\nu}.
\end{align}
This modification accounts for the fact that the half quasi-hole only ``sees'' half of the liquid particles. With this modification, though, the mean-field formula describes well the behavior of impurities in bosonic or fermionic Moore-Read liquids, as is shown in the right panel of Fig.~\ref{MR}. 

In summary, the impurity angular momentum allows to distinguish between different types of quasiholes in the Moore-Read liquid, and to characterize their effective single-particle levels. In future work, we will explore the scenario of multiple impurities in Moore-Read liquids. Particular interest shall be given to the case of a liquid with (at least) four impurities bound to half quasiholes. In this case, the non-Abelian nature of the liquid is reflected by its topological degeneracy. 

\section{Impurities bound to quasiparticles}
So far we have only considered impurities which, through repulsive interactions, bind to quasiholes of a quantum Hall liquid. It is interesting to ask what happens if we have impurities which attract the liquid particles. Can they bind to quasiparticle and exhibit a behavior in close analogy to the impurities bound to quasiholes? Due to the more complicated trial wave functions for quasiparticles (as compared to the quasihole wave functions), we have not been able to answer this question definitely by numerical evaluation. However, here we provide a brief perspective on this subject.

To accommodate a quasihole the angular momentum of the liquid has to be increased by $N_a$ quanta of angular momentum, whereas the formation of a quasiparticle requires to reduce the angular momentum by the same amount. This suggests that an impurity particle bound to a quasiparticle shall carry negative angular momentum. Mathematically, this is possible if the Landau level basis of the impurity is the complex conjugate of the Landau levels for the liquid. Physically, this is possible if the coupling to the gauge field (charge $\times$ magnetic field) is reversed.  Indeed, it seems natural that such reversal occurs for an impurity which binds to a quasiparticle: In this case, the impurity shall carry a charge which is opposite to the charge of the liquid particles (e.g. a positively charged trion which binds to negatively charged electronic quasiparticles), and with this, the impurity will be subject to opposite magnetic fluxes. This leads to the following ansatz for a single impurity, bound to a quasiparticle at position  $w$, in the $m$th effective impurity state:
\begin{align}
 \Psi_{m,{\rm qp}}(w) = w^{*m} \prod_{i=1}^{N_a} (\partial_i - w^*) \Psi_{\rm Laughlin},
\end{align}
where the asterisk denotes the conjugate of the complex position. In this wave function, the impurity ``sees'' the liquid particles as conjugated fluxes, in close analogy to the quasihole case. Thus, we expect that the same re-definition of magnetic length applies also to the effective impurity Landau level. This implies that the effective impurity Landau levels for quasiparticles are  the conjugate of those for quasiholes given by Eq.~(1) in the main text, and accordingly, the impurity angular momentum is given by Eq.~(3) of the main text, yet with opposite sign.

\section{Breakdown of the mean-field theory}
There are different scenarios in which the mean-field result (specifically Eq.~(3) of the main text) ceases to describe the behavior of an impurity bound to a quasihole: (i) Eq.~(3) of the main text breaks down for $\nu=1$, as we already mentioned above in the context of a bosonic Moore-Read liquid. (ii) In the limit of large excitation index $m$, the impurity reaches the edge of the system, and deviations from Eq.~(3) of the main text can be observed.  In the following, we discuss these two breakdown scenarios in more detail.

\begin{figure*}[t]
    \centering 
    \includegraphics[width=0.48\textwidth]{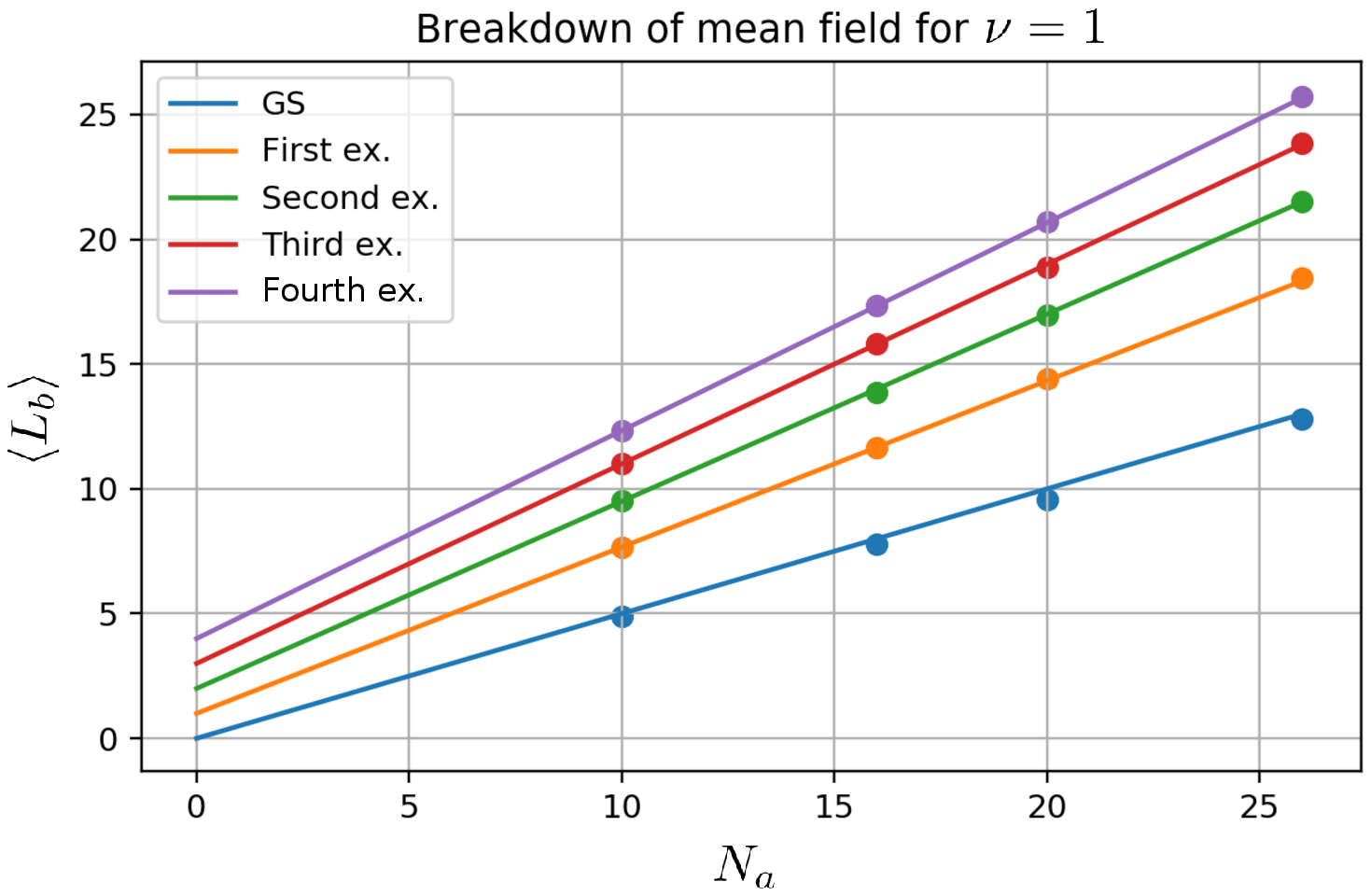}
    \includegraphics[width=0.48\textwidth]{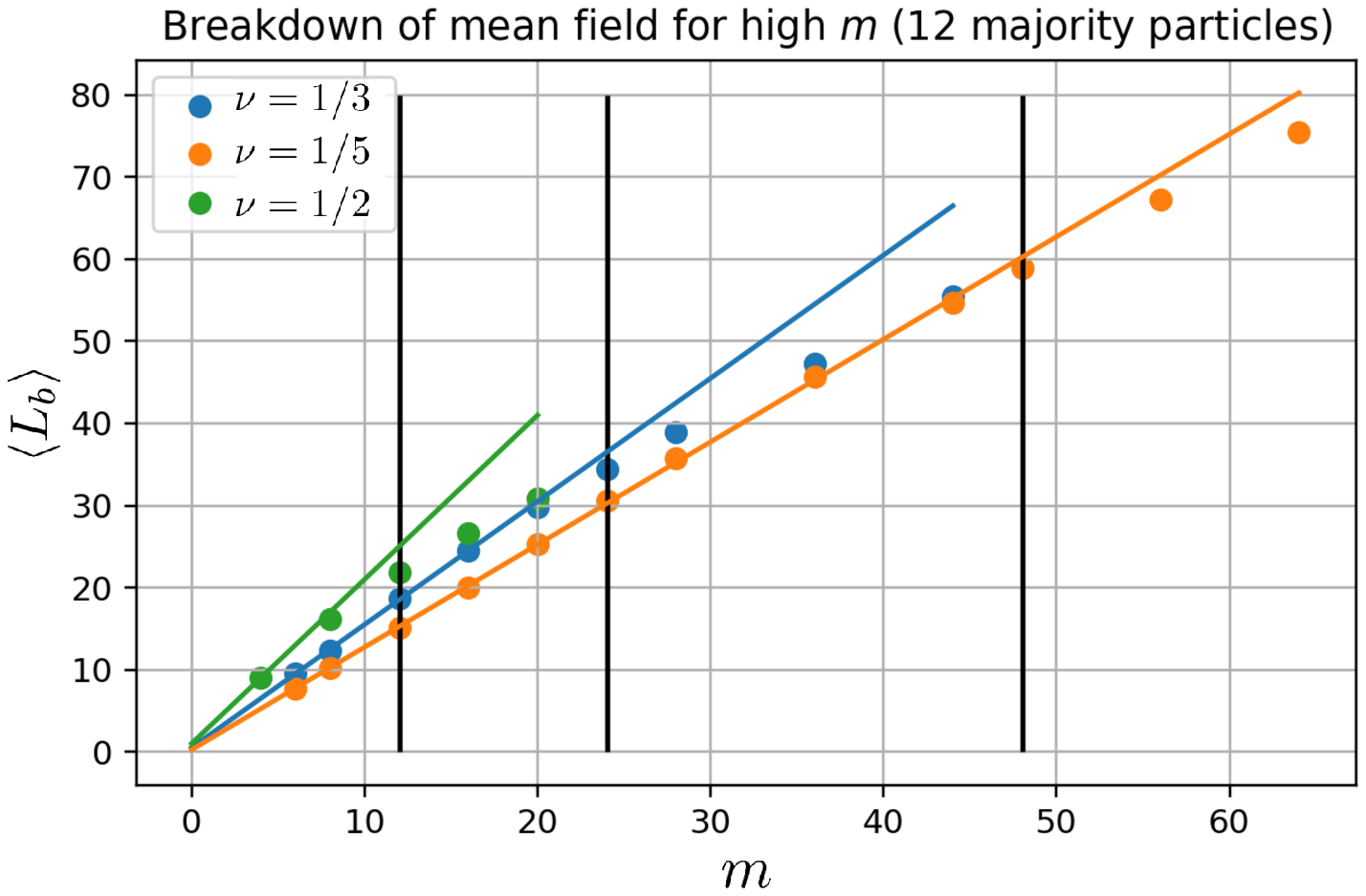}
    \caption{Left: Breakdown of mean field theory at filling $\nu=1$, illustrated by impurity angular momentum vs. $N_a$ for different impurity levels $m$. The impurity angular momentum has become an extensive quantity, and scale according to formula~\ref{ilb} represented by the solid lines. Right: Breakdown of mean field theory at large $m$, illustrated by impurity angular momentum vs. $m$ for three different fillings. The black lines represent the value $(q-1)N$ for the three fillings, at which the impurity should be at the edge of the liquid.}
    \label{BD}
\end{figure*}

\subsection{Breakdown at $\nu=1$}
The mean-field picture, which yields Eq.~(3) of the main text, assumes a screening of the magnetic field due to the liquid particles. In this picture, a liquid at $\nu=1$ would entirely screen the magnetic field, and the assumption of our model that both liquid and impurity are subject to the lowest Landau level breaks down. This restricts Eq.~(3) of the main text to fractional quantum Hall liquids at $\nu<1$.

Thus, it is interesting to ask how an impurity in an integer quantum Hall liquid at $\nu=1$ behaves. Therefore, let us write down the wave function which describes such scenario:
\begin{align}
\label{IQH}
 \Psi_m = w^m \prod_{1\leq i \leq N_a} (w- z_i) \prod_{1\leq i<j\leq N_a} (z_i-z_j).
\end{align}
In the impurity ground state, $m=0$, both the impurity coordinates and the liquid coordinates are on equal footing, thus we can immediately conclude that the impurity angular momentum must be equal to the angular momentum per particle, $L_b^0 = N_a/2$. This establishes a stark contrast to the fractional case: While in the fractional scenario the average angular momentum of the impurity is independent from the number of liquid particles, in the integer case the average impurity angular momentum becomes an extensive quantity. 

We may further ask what happens for excited impurity states, i.e. for $m>0$. To address this scenario, we have computed the average angular momentum of the impurity by Monte Carlo integration applied to the wavefunction in Eq.~(\ref{IQH}). In  the left panel of Fig.~\ref{BD}, the average impurity angular momentum is plotted vs. the number of liquid particles $N_a$, for different $m$. Formally, these results can very well be captured by the following expression:
\begin{align}
\label{ilb}
 L_b^m = N_a \frac{m+1}{m+2}
\end{align}
Thus, the extensive character of $L_b^m$ exists for all $m$. This establishes a clear difference between the fractional and the integer scenario. In particular, in the integer case, $m$ cannot be interpreted as a quasi-quantum number of impurity angular momentum. 

We note that very similar observations as made here for the integer quantum Hall system at $\nu=1$ have also been made for the bosonic Moore-Read liquid at filling $\nu=1$, see also the Section above.

\subsection{Breakdown for large $m$}
The previous subsection demonstrates that, in the fractional quantum Hall liquid at $\nu<1$, the non-extensive nature of the impurity angular momentum is a striking and non-trivial feature. This feature keeps an impurity with small $m$ away from the edge. Nevertheless, it is still possible to make the tracer particle explore the full size of the system, if it is excited to large values of $m$. To determine the value of $m$ at which such breakdown is expected, let us first estimate the size of the liquid: the highest orbital which is occupied in the Laughlin wave function at filling $\nu=1/q$ is $M=q(N_a-1)$, thus the system size is on order of $R \sim \sqrt{2M} l_B = \sqrt{2q(N_a-1)} l_B$. On the other hand, an impurity in level $m$ has a density peak at $W=\sqrt{2 (qm+1)/(q-1)} l_B$. Accordingly, we expect a breakdown of our predictions for $m \approx (q-1)N_a$. 

By explicit Monte Carlo integration, we confirm that the breakdown indeed happens near this expected value of $m$. The results are shown in the right panel of Fig.~\ref{BD}: For a fixed liquid size ($N_a=12$), we plot the average value of the impurity angular momentum as a function of $m$ at different filling fractions $1/q$. The solid lines indicate the behavior expected from the mean-field formula, Eq.~(3) of the main text. The vertical black lines indicate where, according to the above estimate, the deviations from this formula are expected (for the different values of $q$). Indeed, we find that around these values of $m$ the impurity angular momentum remains below the expectations from Eq.~(3) of the main text.

%